\documentclass[10pt]{article}
\usepackage[dvips]{graphicx}

\usepackage{epsfig}
\usepackage{graphicx}
\usepackage{indentfirst}
\usepackage{amsmath}
\usepackage{amsfonts}
\usepackage{amssymb}

\usepackage[pdftex,pdfmenubar=true,bookmarks=true,pdftoolbar=true]{hyperref}
\hypersetup{colorlinks,citecolor=green,filecolor=magenta,linkcolor=red,urlcolor=cyan,pdftex}

\begin{document}

\begin{center}
\begin{flushright}\begin{small}    UFES 2012
\end{small} \end{flushright} \vspace{1.5cm}
\huge{Anisotropic fluid for a set of non-diagonal tetrads in $f(T)$ gravity} 
\end{center}

\begin{center}
{\small  M. Hamani Daouda $^{(a)}$}\footnote{E-mail address:
daoudah8@yahoo.fr}\ ,
{\small  Manuel E. Rodrigues $^{(a)}$}\footnote{E-mail
address: esialg@gmail.com}\ and
{\small    M. J. S. Houndjo $^{(b)(c)}$}\footnote{E-mail address:
sthoundjo@yahoo.fr} \vskip 4mm

(a) \ Universidade Federal do Esp\'{\i}rito Santo \\
Centro de Ci\^{e}ncias
Exatas - Departamento de F\'{\i}sica\\
Av. Fernando Ferrari s/n - Campus de Goiabeiras\\ CEP29075-910 -
Vit\'{o}ria/ES, Brazil \\
(b)  \ Departamento de Engenharia e Ci\^{e}ncias Naturais - CEUNES\\
Universidade Federal do Esp\'irito Santo\\
CEP 29933-415 - S\~ao Mateus - ES, Brazil\\
\ (c) Institut de Math\'{e}matiques et de Sciences Physiques (IMSP)\\01 BP 613 Porto-Novo, B\'{e}nin\\
\vskip 2mm

\end{center}

\begin{abstract}
We consider $f(T)$ gravity for a Weitzenbock's spherically symmetric and static spacetime, where the metric is projected in the tangent space to the manifold, for a set of non-diagonal tetrads. The matter content is coupled through the energy momentum tensor of an anisotropic fluid, generating various classes of new black hole and wormhole solutions. One of these classes is that of cold  black holes. We also perform the reconstruction   scheme of the algebraic function $f(T)$ for two cases where the radial pressure is proportional to $f(T)$ and its first derivative.

\end{abstract}
Pacs numbers: 04.50. Kd, 04.70.Bw, 04.20. Jb


\section{Introduction}
A special attention is now attached to the so-called Teleparallel Theory (TT). This is a geometric theory which possesses only torsion, without curvature, for characterising the gravitational interaction between matter fields. The TT, which is dynamically equivalent to the General Relativity (GR) \cite{pereira}, possesses as fundamental field the tetrads ones with which the Weitzenbock's connection is generated \cite{weitzenbock}. The contraction of the Weitzenbock's connection through a linear combination defines the action of the theory.
\par
With the progress of the measurements about the evolution of the universe, as the expansion and the acceleration, the dark matter and dark energy, various proposals for modifying the  GR are being tested. As the unifications theories, for the scales of low energies, it appears in the effective actions besides the Ricci scalar, the terms  $R^2$, $R^{\mu\nu}R_{\mu\nu}$ and $R^{\mu\nu\alpha\beta}R_{\mu\nu\alpha\beta}$ and the proposal of modified gravity that agrees with the cosmological and astrophysical data is $f(R)$ theory \cite{odintsov,capozziello}. The main problem that one faces with this theory is that the equation of motion is of order 4, being more complicated than the GR for any analysis. Since the GR possesses the TT  as analogous,  it has been thought the so-called $f(T)$  gravity, $T$ being the torsion scalar, which would be the analogous to the generalizes GR, the $f(R)$ gravity. The $f(T)$ gravity is the generalization of the TT as we shall see later. Note also that the $f(T)$ theory is free of the curvature, i.e., it is defined  from the  Weitzenbock's connection. However, it has been shown recently that this theory breaks the invariance of the local Lorentz transformations \cite{barrow}. Other recent problem is that the $f(T)$ gravity appears to be dependent on the used frame, i.e., it is not covariant \cite{barrow, yapiskan}. 
\par
The $f(T)$ gravity  appears in Cosmology as   the source driving the inflation \cite{fiorini}. Also, it has been used as a theory that reproduces the acceleration of the universe without the necessity of introducing the dark energy \cite{ferraro,ratbay,eric}. The contributions of the cosmological perturbations of this theory have been studied \cite{dutta2}. In gravitation, the first model of black hole which has been analysed was the BTZ \cite{fiorini2}. Recently, also in the framework of $f(T)$ gravity, several black hole and wormhole solutions with spherical symmetry have been found \cite{wang,stephane,stephane1}. Moreover, other analysis about various themes are being investigated in the context of $f(T)$ gravity \cite{x}.\par
In this paper, we study some solutions obtained by fixing the spherical symmetries and the staticity  of the metric, as well as the more usual methods of GR \cite{florides}. Introducing an anisotropic matter content, we obtain new black hole and traversable wormhole solution for the  $f(T)$ gravity, considering a set of non-diagonal tetrads.
\par
The paper is organized as follows. In Section $2$, we present a brief revision of the fundamental concepts of the Weitzenbok's geometry, the action of the $f(T)$ gravity and the equations of motion. In Section $3$, we fix the symmetries of the geometry and present the equations of the energy density, the radial and tangential pressures. The Section $4$ is devoted for obtaining new solutions in the $f(T)$ gravity. In  Section $5$, we present a summary of the reconstruction method for the static case of the $f(T)$ theory and reconstruct two simplest cases linked with the radial pressure. The conclusion and perspectives are presented in the Section 6. 
       

\section{\large The field equations from $f(T)$ theory}

The mathematical concept of the $f(T)$ gravity is based on the Weitzenbock's geometry and there exists some excellent works in this way \cite{pereira,pereira2, pereira3}. Our convention and nomenclature are the following: the Latin index describe the elements of the tangent space to the manifold (spacetime), while the Greek ones describe the elements of the spacetime. For a general spacetimes metric, we can define the line element as
\begin{equation}
dS^{2}=g_{\mu\nu}dx^{\mu}dx^{\nu}\; .
\end{equation} 
 This metric can be projected in the tangent space to the manifold, using the representation of the tetrad matrix, where the line element is 
\begin{eqnarray}
dS^{2} &=&g_{\mu\nu}dx^{\mu}dx^{\nu}=\eta_{ij}\theta^{i}\theta^{j}\label{1}\; ,\\
dx^{\mu}& =&e_{i}^{\;\;\mu}\theta^{i}\; , \; \theta^{i}=e^{i}_{\;\;\mu}dx^{\mu}\label{2}\; ,
\end{eqnarray} 
where $\eta_{ij}=diag[1,-1,-1,-1]$ and $e_{i}^{\;\;\mu}e^{i}_{\;\;\nu}=\delta^{\mu}_{\nu}$ or  $e_{i}^{\;\;\mu}e^{j}_{\;\;\mu}=\delta^{j}_{i}$. The square root of the metric determinant is given by  $\sqrt{-g}=\det{\left[e^{i}_{\;\;\mu}\right]}=e$.  Now, we describe the spacetime through the tetrad matrix and then define the Weitzenbock's connection as
\begin{eqnarray}
\Gamma^{\alpha}_{\mu\nu}=e_{i}^{\;\;\alpha}\partial_{\nu}e^{i}_{\;\;\mu}=-e^{i}_{\;\;\mu}\partial_{\nu}e_{i}^{\;\;\alpha}\label{co}\; .
\end{eqnarray}
From the definition of connection (\ref{co}), the spacetime possesses an identically null curvature, having only the torsion contribution and its related quantities for this geometry. Due to the fact that the antisymmetric part of the connection does not vanish, we can define directly from the components of the connection, the torsion tensor, whose components are given by
\begin{eqnarray}
T^{\alpha}_{\;\;\mu\nu}&=&\Gamma^{\alpha}_{\nu\mu}-\Gamma^{\alpha}_{\mu\nu}=e_{i}^{\;\;\alpha}\left(\partial_{\mu} e^{i}_{\;\;\nu}-\partial_{\nu} e^{i}_{\;\;\mu}\right)\label{tor}\;.
\end{eqnarray}
Through the torsion tensor, we can define two important tensors in this geometry: the contorsion and the tensor $S$, whose components can be written as  
\begin{eqnarray}
K^{\mu\nu}_{\;\;\;\;\alpha}&=&-\frac{1}{2}\left(T^{\mu\nu}_{\;\;\;\;\alpha}-T^{\nu\mu}_{\;\;\;\;\alpha}-T_{\alpha}^{\;\;\mu\nu}\right)\label{cont}\; ,\\
S_{\alpha}^{\;\;\mu\nu}&=&\frac{1}{2}\left( K_{\;\;\;\;\alpha}^{\mu\nu}+\delta^{\mu}_{\alpha}T^{\beta\nu}_{\;\;\;\;\beta}-\delta^{\nu}_{\alpha}T^{\beta\mu}_{\;\;\;\;\beta}\right)\label{s}\;.
\end{eqnarray}
We are now able to define easily  the scalar that makes up the action of  $f(T)$ gravity, i.e. the torsion scalar $T$.  Through  (\ref{tor})-(\ref{s}), we define the scalar torsion scalar as  
\begin{eqnarray}
T=T^{\alpha}_{\;\;\mu\nu}S^{\;\;\mu\nu}_{\alpha}\label{tore}\; .
\end{eqnarray}
Following the same idea of the GR and $f(R)$ theory for the coupling of the geometry with the matter part, we define the action of the $f(T)$ gravity as
\begin{eqnarray}
S[e^{i}_{\mu},\Phi_{A}]=\int\; d^{4}x\;e\left[\frac{1}{16\pi}f(T)+\mathcal{L}_{Matter}\left(\Phi_{A}\right)\right]\label{action}\; ,
\end{eqnarray}
where we used the units $G=c=1$ and the $\Phi_{A}$ are the matter fields. Considering the action (\ref{action}) as a functional of the fields  $e^{i}_{\mu}$ and $\Phi_{A}$,  and vanishing the variation of the functional  with respect to the field  $e^{i}_{\nu}$, i.e. the principle of minimum action, one obtains the following equation of motion  \cite{barrow}
\begin{eqnarray}
S^{\;\;\nu\rho}_{\mu}\partial_{\rho}Tf_{TT}+\left[e^{-1}e^{i}_{\mu}\partial_{\rho}\left(ee^{\;\;\alpha}_{i}S^{\;\;\nu\rho}_{\alpha}\right)+T^{\alpha}_{\;\;\lambda\mu}S^{\;\;\nu\lambda}_{\alpha}\right]f_{T}+\frac{1}{4}\delta^{\nu}_{\mu}f=4\pi\mathcal{T}^{\nu}_{\mu}\label{em}\; ,
\end{eqnarray}
where $\mathcal{T}^{\nu}_{\mu}$ is the energy momentum tensor, $f_{T}=d f(T)/d T$ and $f_{TT}=d^{2} f(T)/dT^{2}$. If we consider $f(T)=a_{1}T+a_{0}$, the TT is recovered with a cosmological constant.
\par
We now introduce the matter content as being described by an anisotropic fluid, whose energy-momentum components are given by 
\begin{eqnarray}
\mathcal{T}^{\,\nu}_{\mu}=\left(\rho+p_t\right)u_{\mu}u^{\nu}-p_t \delta^{\nu}_{\mu}+\left(p_r-p_t\right)v_{\mu}v^{\nu}\label{tme}\; ,
\end{eqnarray}
where $u^{\mu}$ is the four-velocity, $v^{\mu}$ the unitary space-like vector in the radial direction, $\rho$ the energy density, $p_r$ the pressure in the direction of $v^{\mu}$ (radial pressure) and $p_t$  the pressure orthogonal to $v_\mu$ (tangential pressure). Since we are assuming an anisotropic spherically symmetric matter, one has $p_r \neq p_t$, such that their equality corresponds to an isotropic fluid sphere.
\par
In the next section, we will make some considerations for the manifold symmetries in order to obtain simplifications in the equations of motion and the specific solutions of these symmetries.

\section{\large   Spherically symmetric geometry}

We consider from the beginning the tetrads matrix as the fundamental fields of the $f(T)$ theory. Now, in the same way that the frames were constructed for the TT theory in \cite{maluf1} and \cite{maluf2}, our tetrad ansatz is elaborated fixing the degree of freedom as follows  $e_{0}^{\;\;\mu}=u^{\mu}$, where $u^{\mu}$ is the four velocity of an observer in free fall, and $e_{1}^{\;\;\mu},e_{2}^{\;\;\mu}$ and $e_{3}^{\;\;\mu}$ are oriented along the unitary  vectors in the Cartesian directions $x$ , $y$ and $z$, resulting in the matrix
\begin{eqnarray}
\left\{e^{i}_{\;\;\mu}\right\}= \left(\begin{array}{cccc}
e^{a/2}&0&0&0\\
0&e^{b/2}\sin\theta\cos\phi &r \cos\theta\cos\phi &-r \sin\theta\sin\phi\\
0&e^{b/2}\sin\theta\sin\phi &r \cos\theta\sin\phi &r \sin\theta\cos\phi\\
0& e^{b/2}\cos\theta & -r\sin\theta & 0
\end{array}\right)\label{tetra}\; ,
\end{eqnarray}
for spherical and static symmetries. Using the relations (\ref{1}) and (\ref{2}), one can write the components of the metric, through the line element of spherically symmetric and static spacetimes as
\begin{equation}
dS^{2}=e^{a(r)}dt^{2}-e^{b(r)}dr^{2}-r^{2}\left[d\theta^{2}+\sin^{2}\left(\theta\right)d\phi^{2}\right]\label{ele}\; .
\end{equation} 

This choice of tetrad matrices is not unique, because the aim of letting the line element invariant under local Lorentz transformations, is to obtain the form (\ref{1}). Other choices have been performed with diagonal matrices, as in references \cite{stephane,stephane1}. Using  (\ref{tetra}), one can obtain $e=\det{\left[e^{i}_{\;\;\mu}\right]}=e^{(a+b)/2}r^2 \sin\left(\theta\right)$, and with (\ref{co})-(\ref{tore}) we determine the  torsion scalar in terms of  $r$
\begin{eqnarray}
T(r) &=& \frac{2e^{-b}}{r^2}\left(e^{b/2}-1\right)\left(e^{b/2}-1-ra^{\prime}\right)\label{te}\; ,
\end{eqnarray} 
where the prime  ($^{\prime}$) denotes the derivative with respect to  the radial coordinate $r$. One can now re-write the equations of motion (\ref{em}) for an anisotropic fluid as 
\begin{eqnarray}
4\pi\rho &=& \frac{f}{4}-\left( \frac{T}{2}-\frac{1}{r^2}-\frac{e^{-b}}{r^2}\left(1-rb^{\prime}\right)\right)\frac{f_T}{2}-\frac{e^{-b/2}}{r}\left(e^{-b/2}-1\right)\left(f_T\right)^{\prime}\,,\label{dens} \\
4\pi p_{r} &=&  \left(\frac{T}{2}+\frac{e^{-b}}{r^2}\left(1+ra^{\prime}\right)-\frac{1}{r^2}\right)\frac{f_T}{2}-\frac{f}{4}\label{presr}\;, \\
4\pi p_{t} &=& \frac{e^{-b}}{2}\left(\frac{a^{\prime}}{2}+\frac{1}{r}-\frac{e^{b/2}}{r}\right)\left(f_T\right)^{\prime}+ \left[\frac{T}{2}+e^{-b}\left(\frac{a''}{2}+\left(\frac{a'}{4}+\frac{1}{2r}\right) (a^{\prime}-b^{\prime})\right)\right]\frac{f_T}{2}-\frac{f}{4}\label{prest}\;,
\end{eqnarray} 
where $p_{r}$ and $p_{t}$  are the radial and tangential pressures respectively. In the case where there does not exist a set of non-diagonal tetrads, as in \cite{stephane,stephane1}, we get an  off-diagonal equation  (component $\theta-r$) which imposed to the algebraic function $f(T)$ to be a linear function of $T$, or the torsion scalar had to be a constant, with a free $f(T)$. Now, in this present paper, with the choice of a set of non-diagonal tetrads, we do not get such constraint equation, but rather, $f(T)$ may assume an arbitrary functional form. In the next section, we will determine new solutions for the $f(T)$ theory making some consideration about the matter component $p_{r} (r)$.



\section{New solutions for a set of non-diagonal tetrads}

In this section, we will study two simplest cases for a set of non-diagonal tetrads. The first is when the radial pressure $p_r$, in (\ref{presr}), is identically  null. This has been done originally in GR by Florides \cite{florides} and used  later by Boehmer et al \cite{boehmer1}, and in $f(T)$ gravity, the same condition has been used for the cases of diagonal tetrads \cite{stephane,stephane1}. We can classify these two cases:
\begin{enumerate}

\item Considering $p_{r}= 0$ in (\ref{presr}), we get 
\begin{eqnarray}
f(T)=2f_{T}(T)\left\{\frac{T}{2}+\frac{1}{r^2}\left[e^{-b}(1+ra^{\prime})-1\right]\right\}\,.\label{p0}
\end{eqnarray}
We can explain two main cases here: 
\begin{enumerate}

\item Considering  $f(T)=\exp\left[a_1 T\right]=\sum\limits_{n=0}^{\infty}[\left(a_{1}T\right)^n/n!]$ in (\ref{p0}), one obtains 
\begin{equation}
T(r)=\frac{1}{a_1}-\frac{2}{r^2}\left[e^{-b}(1+ra^{\prime})-1\right]\label{t1}\,.
\end{equation}
Choosing the quasi-global coordinate $a(r)=-b(r)$ and equating  (\ref{te}) with (\ref{t1}), one gets
\begin{eqnarray}
e^{a(r)}=e^{-b(r)}=\frac{-3+r(6a_1+r^2)+2\sqrt{3}\sqrt{3a_1^2 r^2+a_1r^4-3a_1 r}}{12a_1 r}\label{sol1}\;.
\end{eqnarray}
This is a black hole solution whose the horizon is given by  $r_H=\sqrt[3]{3}$, for $a_1<0$. The torsion scalar (\ref{t1}) is given by   
\begin{equation}
T(r)=\frac{-6\sqrt{3}a_1^2 r+3(2a_1+r^2)\sqrt{a_1 r (3a_1 r+r^3-3)}-\sqrt{3}a_1(4r^3-3)}{6a_1 r^2\sqrt{a_1 r (3a_1 r+r^3-3)}}\label{t2}\;.
\end{equation}

\item Considering 
\begin{equation}
T(r)=\frac{2}{r^2}\left[e^{-b}(1+ra^{\prime})-1\right]\label{t3}\,,
\end{equation}
the equation (\ref{p0}) yields
\begin{equation}
f(T)=\sqrt{\frac{T}{T_0}}=\sum\limits_{n=0}^{\infty}\frac{(-1)^n(2n!)}{(1-2n)(n!)^2(4^n)}\left(\frac{T-T_0}{T_0}\right)^n\,.\label{f0}
\end{equation}

When we impose 
\begin{eqnarray}
e^{a(r)}=\left(1-\frac{r_H}{r}\right)^k\,,\label{cond2}
\end{eqnarray}
with  $k\geq 2$, equating (\ref{te}) with  (\ref{t3}), we obtain
\begin{eqnarray}
e^{b(r)}=\left[\frac{2r+(k-2)r_H}{2(r-r_H)}\right]^2\,,\,T(r)=\frac{-2k^2r_{H}^2}{r^2[2r+(k-2)r_{H}]^2}\,.\label{b1}
\end{eqnarray}
This is the first class of cold black holes solutions\footnote{Solutions that possess identically null Hawking temperature.} in $f(T)$ gravity. The degenerated event horizon of order $k$ is obtained in $r=r_H$. 
\par
We can still look at the limit where  $T<<T_0$ in (\ref{f0}), which, at the second order, leads to 
\begin{equation}
f(T)=\sqrt{\frac{T}{T_0}}\approx a_0+a_1 T+a_2 T^2\,,
\end{equation}
where $a_0=3/8$, $a_1=3/(4T_0)$ and  $a_2=-1/(8T_0^2)$. In this limit, the theory becomes the TT one plus a quadratic term in the torsion scalar $T$.

\item With the choice 
\begin{eqnarray}
e^{b(r)}=\frac{r_0}{r}\label{cond3}\;,
\end{eqnarray}
and equating  (\ref{te}) with (\ref{t3}), one gets 
\begin{eqnarray}
a(r)=-4\sqrt{\frac{r_0}{r}}-2\ln r\label{a1}\;.
\end{eqnarray}
This solution is completely different from that obtained in  \cite{stephane1}, taking the same condition of coordinate, but for a set of diagonal tetrads.
\end{enumerate}

\item Let us consider here the case where the radial pressure is simply a function of the radial coordinate $r$ and the algebraic function $f(T)$ is chosen as
\begin{eqnarray}
f(T)=a_2 T^2+a_1 T+a_0\label{f1}\;.
\end{eqnarray}
In this case, substituting (\ref{f1}) into (\ref{presr}) and taking the coordinate condition \cite{stephane1}
\begin{eqnarray}
a(r)=\ln\left(\frac{r_0}{r}\right)\;,\label{cond4}
\end{eqnarray}
we obtain 
\begin{eqnarray}
T(r)=\frac{2}{r^2}\left[1\pm\sqrt{1-\frac{a_1}{2a_2}r^2+\frac{r^4}{4a_2}(a_0+16\pi p_{r}(r))}\right]\label{t2}\;.
\end{eqnarray}
Equating (\ref{t2}) with  (\ref{te}) and taking into account  (\ref{cond4}), we get the solution 
\begin{eqnarray}
dS^2=\frac{r_0}{r}dt^2-\left[1-\frac{r^2}{2}T(r)\right]^{-2}dr^2-r^2d\Omega^2\label{sol3}\;,
\end{eqnarray}
where  $T(r)$ is given in (\ref{t2}) and the condition 
\begin{equation}
1-\frac{a_1}{2a_2}r^2+\frac{r^4}{4a_2}[a_0+16\pi p_{r}(r)]\geq0\,,\label{pr0}
\end{equation}
has to be satisfied for $T(r)$ being a real function.  This is a new class of  traversable wormhole solutions. We can observe this by following the same process as in \cite{stephane}. This line element (\ref{ele}) can be put in the form  
\begin{equation}
dS^{2}=e^{a(r)}dt^{2}-dl^{2}-r^{2}(l)d\Omega^{2}\label{elw}\;,
\end{equation}
where $a(r)$ is denoted redshift function, and through the redefinition $\beta (r)=r\left[1-e^{-b(r)}\right]$, with $b(r)$ being the metric function given in (\ref{ele}), $\beta(r)$ is called  shape function. Therefore, the conditions of existence of a traversable wormhole are: a) the function $r(l)$ must possess a minimum value $r_{1}$ for $r$, which imposes ${d^2r}(l)/dl^{2}>0$; b) $\beta(r_{1})=r_{1}$;  c) $a(r_{1})$ has a finite value; and finally d) $d\beta(r)/dr|_{r=r_{1}}\leqslant  1$.
\par
In this case, $\beta(r)=r[1-(1-r^2T(r)/2)^2]$. Hence, the conditions of a traversable wormhole are applied on the torsion scalar $T(r)$, in (\ref{t2}). The condition a), which can be explained in $d^2r/dl^2=[\beta (r)-r\beta^{\prime}(r)]/2r^2=-r[1-r^2T(r)/2][T(r)+rT^{\prime}(r)/2]>0$, leads to  $[T(r)+rT^{\prime}(r)/2]<0$; the condition b), yields $T(r_1)=2/r_1^2$; the condition c) is always satisfied, as we can see through (\ref{sol3}); and finally, the condition  d) leads to $[1-r^2T(r)/2]\geq 2r^2[T(r)+rT^{\prime}(r)/2]$, which is always true when  a) is satisfied.
\par
Therefore, we must choose the torsion scalar, which will be defined from the radial pressure $p_r(r)$, in (\ref{t2}), such that it satisfies the conditions a)-d) and (\ref{pr0}), for obtaining traversable wormhole solutions. Let us consider first a class of solutions coming from the following example, $T(r)=2r_1^{m-2}/r^m$, with $r_1>0$ and $m\geq 3$, which is obtained for $p_r (r)=[2a_1 r^{2m}r_1^{4}-a_0 r^{2(m+1)}r_1^4-8a_2r^{m}r_1^{m+2}+4a_2 r^{2}r_1^{2m}]/16\pi r_1^4 r^{2(m+1)}$, which satisfies (\ref{pr0}). The condition  a) is satisfied, since $d^2 r/dl^2=(r_1^{m-2}/r^{m-1})[1-(r_1/r)^{m-2}]>0$; the conditions b) and c) are also directly satisfied; and d) is satisfied through a).    
\end{enumerate}
\section{Reconstruction in static f(T) theory}
A method widely used in cosmology in order to obtain the algebraic form of the gravitational part of the action is the so-called reconstruction scheme. This method stems from the introduction of an auxiliary field for the reconstruction of the algebraic function of the main action, as in the case of the theory $f(R)$ for example \cite{reconstruction1}.  
\par
We can briefly present this method as follows. Considering the algebraic function
\begin{equation}
f(T)=P\left(\varphi\right)T+Q\left(\varphi\right)\label{fr}\;,
\end{equation} 
the functional variation of the action  (\ref{action}), with respect to  $\varphi$, is given by 
\begin{equation}
\frac{\delta S}{\delta \varphi}=\frac{e}{16\pi}\left[\frac{dP}{d \varphi}T+\frac{d Q}{d\varphi}\right]=0\label{econst}\;.
\end{equation}  
Solving this equation, we get $\varphi\equiv\varphi(T)$, then,  $f(T)=P[\varphi(T)]T+Q[\varphi(T)]$. Hence, we have the following identities
\begin{eqnarray}
f_{T}(T)&=&P+\left(\frac{dP}{d\varphi}T+\frac{dQ}{d\varphi}\right)\frac{d\varphi}{dT}=P[\varphi(T)]\label{ftr}\;,\\
f_{TT}(T)&=&\frac{dP[\varphi(T)]}{dT}\label{fttr}\;.
\end{eqnarray} 
Having in hand the equations (\ref{fr}), (\ref{ftr}) and (\ref{fttr}), and substituting into (\ref{dens})-(\ref{prest}), we get 
\begin{eqnarray}
4\pi\rho&=&-\frac{e^{-b/2}}{r}\left(e^{-b/2}-1\right)\frac{dP}{dT}+\frac{Q}{4}+\frac{P}{2r^2}\left[1+e^{-b}(1-rb^{\prime})\right]\;,\label{densr}\\
4\pi p_{r}&=&\frac{P}{2r^2}\left[-1+e^{-b}(1+ra^{\prime})\right]-\frac{Q}{4}\label{presrr}\;,\\
4\pi p_{t}&=& \frac{e^{-b}}{2}\left(\frac{a^{\prime}}{2}+\frac{1}{r}-\frac{e^{-b/2}}{r}\right)\frac{dP}{dT}+P\frac{e^{-b}}{4}\left[a^{\prime\prime}+\left(\frac{1}{r}+\frac{a^{\prime}}{2}\right)(a^{\prime}-b^{\prime})\right]-\frac{Q}{4}\label{prestr}\;.
\end{eqnarray}
Since $\varphi$ is an arbitrary field, the reconstruction can be performed directly choosing $\varphi=r$. This method may be used for re-obtaining or reconstructing, when the inversion $r\equiv r(T)$ is possible, the $f(T)$ theory in the static case.
\par
Let us draw up to cases here:
\begin{enumerate}

\item For the case where $f(T)$ is a linear algebraic function, one has the condition $P\in\Re$, due to  (\ref{fttr}). Considering now the condition (\ref{cond4}), and that the radial pressure (\ref{presrr}) obeys
\begin{eqnarray}
p_{r}=g_1 f(T)\label{presr2}\;,
\end{eqnarray}
where $g_1\in\Re$, we get
\begin{eqnarray}
Q&=&-\left(\frac{2}{16\pi g_1 +1}\right)\frac{P}{r^2}-\left(\frac{16\pi g_1}{16\pi g_1+1}\right)PT\,,\\
f(T)&=&\left(\frac{1}{16\pi g_1+1}\right)PT-\left(\frac{2}{16\pi g_1+1}\right)\frac{P}{r^2}\;.\label{fr1}
\end{eqnarray} 
Differentiating  (\ref{fr1}) with respect to  $T$, equating it to  $P$ and integrating, we obtain the relation 
\begin{eqnarray}
r(T)=\left[8\pi g_1(T_0-T)\right]^{-1/2}\;,\label{r1}
\end{eqnarray}
where $T_0$ is an integration real constant. Substituting (\ref{r1}) in (\ref{fr1}), the algebraic function $f(T)$ is then constructed as 
\begin{eqnarray}
f(T)=PT+\left(\frac{-16\pi g_1 PT_0}{16\pi g_1+1}\right)\;.\label{fr2}
\end{eqnarray}
We see that $Q$ is also a constant, given by the second term of (\ref{fr2}).  Substituting  (\ref{te}) into (\ref{r1}), one gets the  following solution 
\begin{eqnarray}
dS^{2}=\frac{r_0}{r}dt^2-\left(\frac{16\pi g_1+1}{16\pi g_1}-\frac{T_0}{2}r^2\right)^{-2}dr^2-r^2d\Omega^2\label{sol6}\;.
\end{eqnarray} 
This is a new traversable wormhole solution. Following the same process of the solution (\ref{sol3}), we obtain  $\beta(r)=r[1-(k-T_0 r^2/2)^2]$, with $k=(16\pi g_1+1)/16\pi g_1$, and then, the minimum of  $r$ is given by $r_1=\sqrt{2k/T_0}$. The condition a), which in this case is given by $d^2 r/dl^2=-T_0 r(k-T_0 r^2/2)$, is satisfied for $T_0<0$ and $-1/16\pi <g_1<0$. The conditions  b) and  c) are directly satisfied, and  d) is also satisfied for $\beta^{\prime}(r_1)=1$.   

\item Taking the condition (\ref{cond4}) and considering that the radial pressure obeys \cite{stephane1}
\begin{eqnarray}
4\pi p_r=g_1f(T)+g_2f_T (T)\label{presr3}\;,
\end{eqnarray}
where $g_1,g_2\in\Re$, we get from (\ref{presrr}) that
\begin{eqnarray}
Q=-\frac{4P}{4g_1+1}\left(g_1 T+g_2+\frac{1}{r^2}\right)\;,\;f(T)=\frac{4P}{4g_1+1}\left(\frac{T}{4}-g_2-\frac{1}{2r^2}\right)\label{fr3}\;.
\end{eqnarray}
Differentiating (\ref{fr3}) with respect to $T$ and equating it to  $P$, we get 
\begin{eqnarray}
g_1 P=\frac{dP}{dT}\left(\frac{T}{4}-g_2-\frac{1}{2r^2}\right)+\frac{P}{r^3}\frac{dr}{dT}\label{p1}\;.
\end{eqnarray}
In order to integrate this equation, we consider first  $P(r)=r^k$, then, (\ref{p1}) becomes 
\begin{eqnarray}
g_1 r=\left[\frac{k}{4}(T-4g_2)+\frac{1}{r^2}\left(1-\frac{k}{2}\right)\right]\frac{dr}{dT}\label{p2}\;,
\end{eqnarray}
which is separable only for $k=2$, yielding 
\begin{eqnarray}
r(T)=r_1\left(\frac{T-4g_2}{T_0-4g_2}\right)^{2g_1}\label{r3}\;,
\end{eqnarray} 
where  $T_0$ and  $r_1$ are integration real constants. Substituting (\ref{r3}) into (\ref{fr3}), the algebraic function $f(T)$ in then reconstructed as
\begin{eqnarray}
f(T)&=&\frac{1}{4g_1+1}\left[r_1^2\frac{(T-4g_2)^{4g_1+1}}{(T_0-4g_2)^{4g_1}}-2\right]\label{fr4}\;.
\end{eqnarray}
This algebraic polynomial function $f(T)$ can be better extended for a particular more simplest case. One can illustrate the particular case in which $4g_1+1=2$, which leads to 
\begin{eqnarray}
f(T)=a_1 T+\left(a_2 T^2+a_0\right)\label{fr5}\;,
\end{eqnarray}
where $a_2=r_1^2/2(T_0-4g_2)\,,\,a_1=-4g_2 r_1^2/(T_0-4g_2)$ and  $a_0=-1$. Then, we see that our reconstruction scheme leads to a function which presents the contributions of higher orders in the torsion scalar, added to the TT term, which in this case is of second and zero orders in (\ref{fr5}).                
\end{enumerate}

\section{Conclusions}
The particular choice of the representation of  a spherically symmetric and static metric of Weitzenbock's spacetime can be done in an arbitrary way, when we wish to maintain the invariance under local Lorentz transformations. Therefore, when we choose a set of diagonal tetrads, for a projection in the tangent space to the manifold,  the imposition of the function $f(T)$ being linear in $T$, or a constant torsion scalar, is recovered. But, as described in this work, choosing a set of non-diagonal tetrads, we can get an arbitrary algebraic function $f(T)$. Therewith, the second derivative of the function $f(T)$ appears in the contribution of the energy density and the tangential pressure in (\ref{dens}) and (\ref{prest}).
\par
We shown some examples of new black holes and wormholes solutions. Through the fixation of the null radial pressure and the coordinates conditions $a(r)=-b(r)$ and (\ref{cond3}), we illustrated an example of a black hole, (\ref{sol1}), of a cold black hole, in (\ref{b1}), and a solution for comparing  the same conditions, with the diagonal case, in (\ref{a1}).
\par
We also imposed to the radial pressure to be just function of the radial coordinate, and $f(T)$ to be a polynomial function of second order in $T$. We then obtained, through the coordinate condition (\ref{cond4}), a new class of traversable wormhole solutions in (\ref{sol3}).
\par
Finally, we made a summary of the reconstruction scheme for our static case of $f(T)$ gravity. For a particular case of the algebraic function $f(T)$ being a linear function, we were able to reconstruct it, considering that it is proportional the radial pressure. This case is that of a new solution of traversable wormhole. Taking into account the condition in which the radial pressure is a linear combination  of the algebraic function $f(T)$ and its derivative, we reconstructed $f(T)$ as a polynomial function of the torsion scalar $T$, where we shown the particular case in which $f(T)$ gravity would be the TT plus some terms of second and zero orders of the torsion scalar $T$.

\vspace{0,25cm}
{\bf Acknowledgement:}   M. H. Daouda thanks CNPq/TWAS for financial support. M. E. Rodrigues  thanks  UFES for the hospitality during the development of this work. M. J. S.  Houndjo thanks CNPq/FAPES for financial support.


\begin{thebibliography}{10}

\bibitem{weitzenbock}R. Weitzenbock, Noordhoff, Groningen; Chap. XIII, Sec 7  (1923).
\bibitem{pereira}R. Aldrovandi and J. G. Pereira,  An Introduction to Teleparallel Gravity, Instituto de Fisica Teorica, UNSEP, Sao Paulo (http://www.ift.unesp.br/gcg/tele.pdf).
\bibitem{odintsov} S. Nojiri and S. D. Odintsov, Phys. Rept. {\bf 505}, 59-144 (2011); S. Nojiri and S.D. Odintsov, 	ECONF C {\bf 0602061}:06, (2006); Int. J. Geom. Meth. Mod. Phys.{\bf 4}: 115-146, (2007).
\bibitem{capozziello}S. Capozziello and V. Faraoni, Beyond Einstein Gravity, A Survey of Gravitational Theories for Cosmology and Astrophysics, Series: Fundamental Theories of Physics, Vol. 170, Springer, New York (2011).
\bibitem{barrow}Baojiu Li, T. P. Sotiriou, and J. D. Barrow, Phys. Rev. D {\bf 83}, 064035 (2011); Phys.Rev.D {\bf 83}:104030 (2011).
\bibitem{yapiskan} Cemsinan Deliduman and Baris Yapiskan, Absence of Relativistic Stars in f(T) Gravity, arXiv:1103.2225v3 [gr-qc].
\bibitem{fiorini}Rafael Ferraro and Franco Fiorini, Phys.Rev. D {\bf75}, 084031 (2007).
\bibitem{ferraro}G. Bengochea and R. Ferraro, Phys.Rev. D {\bf79}:124019 (2009).
\bibitem{ratbay}Ratbay Myrzakulov, Accelerating universe from F(T) gravities, arXiv:1006.1120v1 [gr-qc].
\bibitem{eric}Eric V. Linder, Phys.Rev. D {\bf 81}:127301 (2010).
\bibitem{dutta2}Baojiu Li, Thomas P. Sotiriou, John D. Barrow, Phys.Rev.D {\bf 83}:104017 (2011); Shih-Hung Chen, J. B. Dent, S. Dutta and E. N. Saridakis, Phys.Rev.D {\bf 83}:023508 (2011).
\bibitem{fiorini2}Rafael Ferraro and Franco Fiorini, Phys.Rev. D {\bf 78}:124019 (2008). 
\bibitem{wang}Tower Wang, Phys.Rev. D {\bf 84}:024042 (2011).  
\bibitem{stephane}M. Hamani Daouda, Manuel E. Rodrigues and M. J. S. Houndjo, Eur.Phys.J. C {\bf 71}: 1817 (2011); arXiv:1108.2920v4 [astro-ph.CO].
\bibitem{stephane1}M. Hamani Daouda, Manuel E. Rodrigues and M. J. S. Houndjo, Static Anisotropic Solutions in f(T) Theory,  Eur. Phys. J. C {\bf 72} (2012) 1890; arXiv:1109.0528v3 [physics.gen-ph].
\bibitem{florides}P. S. Florides, A new interior Schwarzschild solution, Proc. R. Soc. Lond. A {\bf337}, 529-535 (1974).
\bibitem{pereira2}R. Aldrovandi, J. G. Pereira and K. H. Vu, Brazilian Journal of Physics, vol. {\bf 34}, no. 4A, December (2004).
\bibitem{pereira3}H. I. Arcos and J. G. Pereira, Int.J.Mod.Phys. D{\bf 13}: 2193-2240 (2004).
\bibitem{boehmer1}Christian G. Boehmer and Francisco S. N. Lobo, Int.J.Mod.Phys. D {\bf 17}:897-910 (2008).
\bibitem{reconstruction1} S. Nojiri, S. D. Odintsov, Phys. Rev. D {\bf 74}: 086005 (2006).
\bibitem{x}Jie Yang, Yun-Liang Li, Yuan Zhong and Yang Li, arXiv:1202.0129v1 [hep-th]; K. Karami and A. Abdolmaleki, arXiv:1201.2511v1 [gr-qc]; K. Atazadeh and F. Darabi, arXiv:1112.2824v1 [physics.gen-ph]; Hao Wei, Xiao-Jiao Guo and Long-Fei Wang, Phys.Lett.B {\bf 707}:298-304 (2012); K. Karami, A. Abdolmaleki, arXiv:1111.7269v1 [gr-qc]; P.A. Gonzalez, Emmanuel N. Saridakis  and Yerko Vasquez, arXiv:1110.4024v1 [gr-qc]; S. Capozziello, V. F. Cardone, H. Farajollahi and A. Ravanpak, Phys.Rev.D {\bf 84}:043527 (2011); Rong-Xin Miao, Miao Li and Yan-Gang Miao, arXiv:1107.0515v3 [hep-th]; Xin-he Meng and Ying-bin Wang, Eur.Phys.J. C {\bf 71}:  1755 (2011); Hao Wei, Xiao-Peng Ma and Hao-Yu Qi, Phys.Lett.B {\bf 703}:74-80 (2011); Miao Li, Rong-Xin Miao and Yan-Gang Miao, JHEP {\bf 1107}:108 (2011); Surajit Chattopadhyay and Ujjal Debnath, Int.J.Mod.Phys.D {\bf 20}:1135-1152 (2011); Piyali Bagchi Khatua, Shuvendu Chakraborty and Ujjal Debnath, arXiv:1105.3393v1 [physics.gen-ph]; Yi-Fu Cai, Shih-Hung Chen, James B. Dent, Sourish Dutta and Emmanuel N. Saridakis, Class. Quantum Grav. {\bf 28}:  215011 (2011); Rong-Jia Yang, Europhys.Lett. {\bf 93}:60001 (2011); Christian G. Boehmer, Atifah Mussa and Nicola Tamanini, Class.Quant.Grav. {\bf 28}: 245020 (2011).

\bibitem{maluf1}J.W. Maluf and S.C. Ulhoa, Gen.Rel.Grav. 41 (2009) 1233-1247; arXiv:0810.1934 [gr-qc].

\bibitem{maluf2}J.W. Maluf, F.F. Faria and S.C. Ulhoa, Class.Quant.Grav. {\bf 24}:  2743-2754 (2007);  arXiv:0704.0986 [gr-qc]; J.W. Maluf, S. C. Ulhoa and J. F. da Rocha-Neto, Phys. Rev. D {\bf 85}: 044050 (2012).
\end{thebibliography}
\end{document}